\documentclass[preprint, superscriptaddress, showpacs,preprintnumbers,amsmath,amssymb,pra]{revtex4}
\usepackage{graphicx}

\begin{document}

\thispagestyle{empty}

\title{Classical Casimir-Polder force between polarizable
microparticles and thin films
including graphene}

\author{
G.~L.~Klimchitskaya}
\affiliation{Central Astronomical Observatory
at Pulkovo of the Russian Academy of Sciences,
St.Petersburg, 196140, Russia}
\affiliation{Institute of Physics, Nanotechnology and
Telecommunications, St.Petersburg State
Polytechnical University, St.Petersburg, 195251, Russia}
\author{
 V.~M.~Mostepanenko}
\affiliation{Central Astronomical Observatory
at Pulkovo of the Russian Academy of Sciences,
St.Petersburg, 196140, Russia}
\affiliation{Institute of Physics, Nanotechnology and
Telecommunications, St.Petersburg State
Polytechnical University, St.Petersburg, 195251, Russia}

\begin{abstract}
We derive analytic expressions for the classical Casimir-Polder
free energy and force for a polarizable (magnetizable) atom
(microparticle) interacting with thin films, made of different
materials, or graphene. It is shown that for an isolated
dielectric film the free energy and force  decrease quicker with
separation, as compared to the case of atom interacting with
a thick plate (semispace). For metallic films some peculiar
features depending on the model of a metal used are analyzed.
For an atom interacting with graphene we demonstrate that at
room temperature the classical regime is achieved at about
$1.5\,\mu$m separation. In this regime the contributions to
the free energy and force due to atomic magnetic polarizability
are suppressed, as compared to main terms caused by the atomic
electric polarizability. According to our results, at
separations above $5\,\mu$m the Casimir-Polder interaction of
atoms with graphene is of the same strength as with an
ideal-metal plane. The classical interaction of atoms with thin films
deposited on substrates is also considered.
\end{abstract}
\pacs{34.35.+a, 37.30.+i, 12.20.Ds, 78.67.Wj}

\maketitle
\section{Introduction}

The Casimir-Polder interaction between polarizable atoms or other
microparticles and a cavity wall has long been investigated
\cite{1,2,3}. A renewed interest to this subject was generated by
the experiments on quantum reflection \cite{4,5,6} and
Bose-Einstein condensation \cite{7,8,9}.
As a result, the dependence of the Casimir-Polder force on atomic
characteristics and material properties of the wall was
investigated in detail \cite{10,11,12,13,14,15} on the basis of
the Lifshitz theory \cite{2}.
Measurements of the thermal Casimir-Polder force between a
condensate of ${}^{87}$Rb atoms and a SiO${}_2$ plate \cite{15a}
were used to obtain constraints on the Yukawa-type corrections
to Newtonian gravitational law in the micrometer interaction
range \cite{15b}.
 Much attention was paid also to
atom-wall interaction for the case of electric and magnetic
polarizable (i.e., magnetizable) atom near a magnetodielectric
wall \cite{16,17,18}. In parallel with atom-wall interaction,
there was an increasing interest to the interaction of
nanoparticles with material surfaces \cite{19,20,21,22}.
Specifically, it was shown \cite{20} that interaction between
a nanosphere of radius $1\,\mbox{nm}\leq R\leq 20\,$nm
and a metallic plane spaced at a separation $a$ is well
described by the Lifshitz formula for atom-wall interaction
under a condition $a\geq 100R$. Keeping in mind that nonmagnetic
metallic nanoparticles can have relatively large magnetic
polarizability due to eddy currents \cite{23}, the respective
Casimir-Polder force can be much larger than in the case of
atoms.

With discovery of graphene and other carbon nanostructures
(see reviews in Refs.~\cite{24,25}) a lot of attention has been
paid to the interaction of these nanostructures with atoms, molecules
and other microparticles \cite{26,27,28,29,30,31,32,33,34,35,35a}.
This was motivated by both fundamental interest and prospective
applications, e.g., to the problem of hydrogen storage \cite{31,34}.
After developing the Dirac model of graphene (see Ref.~\cite{24}
for a review), it was applied to calculate the van der Waals
and Casimir interactions between two graphene sheets and between
a graphene and a plate made of ordinary materials
\cite{36,37,38,39,40,41}.
The obtained reflection coefficients of electromagnetic
oscillations
on graphene were used to calculate the Casimir-Polder interaction
of graphene with different atoms \cite{42,43,44,45}.
In so doing the two theoretical approaches have been elaborated.
In the framework of one approach the reflection coefficients are
expressed in terms of the components of the polarization tensor
in (2+1)-dimensional space-time \cite{36,38,40,41,42,43,44}.
In the framework of another approach they are
effectively expressed in terms
of the density-density correlation functions \cite{37,39,45}.

In this paper we find the classical limit of the thermal
Casimir-Polder force acting between
a polarizable (magnetizable) atom or a
microparticle and a thin film made of various materials
(isolated or deposited on a substrate). We also replace a thin
material film of some thickness with a two-dimensional graphene
sheet described using the Drude model and compare the results
obtained. The consideration of the classical limit (i.e., the
case of sufficiently large separations or high temperatures when
the main contribution to the force does not depend on the
Planck constant \cite{46}) allows to obtain all the results in
a straightforward analytic form. This simplifies the comparison
and helps to reveal the physical role of film thickness.
We consider both magnetodielectric and metallic films and find
that the dominant contribution to the free energy and force
for graphene is in agreement with the case of nonmagnetic
metallic films. It is shown that the magnetic properties of
a microparticle have no impact on the Casimir-Polder
interaction with graphene.

The paper is organized as follows.  In Sec.~II we introduce the
general formalism and consider the classical Casimir-Polder
interaction of an electrically and magnetically
polarizable microparticle
with an isolated magnetodielectric film. In Sec.~III the same
problem is solved for the case of metallic film characterized
by the dielectric permittivity $\varepsilon(\omega)$ and
magnetic permeability $\mu(\omega)$. Section~IV is devoted
to the classical Casimir-Polder interaction of a microparticle
with a graphene sheet. In Sec.~V we consider
the classical Casimir-Polder interaction of a microparticle with
dielectric films deposited on either magnetodielectric or
metallic substrates. In Sec.~VI the reader will find our
conclusions and discussion.

\section{Microparticle interacting with a magnetodielectric
film}

For use in this and following sections, we consider a ground
state atom or a microparticle in vacuum characterized by the
electric polarizability $\alpha(\omega)$ and the magnetic
polarizability $\beta(\omega)$ at a distance $a$ above a sample
consisting of thin film of thickness $D$ deposited on a
thick plate (semispace). Both the film and the semispace are
characterized by the dielectric permittivities
$\varepsilon^{(1,2)}(\omega)$ and magnetic permeabilities
$\mu^{(1,2)}(\omega)$, respectively (see Fig.~1). They
might be made of either dielectric or metallic materials.
The Casimir-Polder interaction between an atom (microparticle)
and a sample is described by the Lifshitz theory \cite{2}.
At relatively large separations or, equivalently, high
temperatures satisfying the conditions
\begin{equation}
a\gg a_T\equiv\frac{\hbar c}{2k_BT},\qquad
T\gg T_{\rm eff}\equiv\frac{\hbar c}{2ak_B},
\label{eq1}
\end{equation}
\noindent
where $k_B$ is the Boltzmann constant, the so-called
{\it classical limit} \cite{46} holds, where up to
exponentially small corrections the Casimir and Casimir-Polder
free energy and force are given by the zero-frequency
contributions to respective Lifshitz formulas \cite{47}.
For ordinary materials at room temperature
$a_T\approx 3.8\,\mu$m, and the classical limit is already
achieved at separations above approximately $5\,\mu$m
(for graphene the classical limit starts from much shorter
separations, see below).

We choose the coordinate plane $(x,y)$ coinciding with the
upper film surface and the $z$ axis perpendicular to it
(see Fig.~1). Then, the Casimir-Polder free energy with account
of magnetic properties of a microparticle, a film and a plate
in the classical limit is given by \cite{16,17,18,47}
\begin{eqnarray}
&&
{\cal F}(a,T)=-k_BT\int_{0}^{\infty}\!\!k_{\bot}^2dk_{\bot}
e^{-2ak_{\bot}}
\label{eq2} \\
&&~~~~~~~~~~\times
[\alpha(0)R_{\rm TM}(0,k_{\bot})+
\beta(0)R_{\rm TE}(0,k_{\bot})].
\nonumber
\end{eqnarray}
\noindent
Here, $R_{\rm TM,TE}$ are the reflection coefficients of the
electromagnetic fluctuations on our sample for two independent
polarizations, transverse magnetic (TM) and transverse electric
(TE), $\mbox{\boldmath$k$}_{\bot}$ is the projection of the
wave vector on the plane $(x,y)$, and
$k_{\bot}=|\mbox{\boldmath$k$}_{\bot}|$.
The explicit expressions for these reflection coefficients are
given by \cite{14,47,48}
\begin{equation}
R_{\rm TM,TE}(0,k_{\bot})=\frac{r_{\rm TM,TE}^{(0,1)}(0,k_{\bot})
+r_{\rm TM,TE}^{(1,2)}(0,k_{\bot})\,
e^{-2k^{(1)}(0,k_{\bot})D}}{1+r_{\rm TM,TE}^{(0,1)}(0,k_{\bot})
r_{\rm TM,TE}^{(1,2)}(0,k_{\bot})\,
e^{-2k^{(1)}(0,k_{\bot})D}}.
\label{eq3}
\end{equation}
\noindent
Here, the Fresnel coefficients $r_{\rm TM,TE}^{(n,n^{\prime})}$
describe reflection on the boundary planes between the vacuum and
the semispace made of the film material ($n=0,\,n^{\prime}=1$) and
between the semispace made of the film material and the thick plate
($n=1,\,n^{\prime}=2$). They are calculated along the imaginary
frequency axis
\begin{eqnarray}
&&
r_{\rm TM}^{(n,n^{\prime})}(i\xi,k_{\bot})=
\frac{\varepsilon^{(n^{\prime})}(i\xi)k^{(n)}(i\xi,k_{\bot})-
\varepsilon^{(n)}(i\xi)
k^{(n^{\prime})}(i\xi,k_{\bot})}{\varepsilon^{(n^{\prime})}(i\xi)
k^{(n)}(i\xi,k_{\bot})+\varepsilon^{(n)}(i\xi)
k^{(n^{\prime})}(i\xi,k_{\bot})},
\nonumber \\
&&
r_{\rm TE}^{(n,n^{\prime})}(i\xi,k_{\bot})=
\frac{\mu^{(n^{\prime})}(i\xi)k^{(n)}(i\xi,k_{\bot})-
\mu^{(n)}(i\xi)
k^{(n^{\prime})}(i\xi,k_{\bot})}{\mu^{(n^{\prime})}(i\xi)
k^{(n)}(i\xi,k_{\bot})+\mu^{(n)}(i\xi)
k^{(n^{\prime})}(i\xi,k_{\bot})},
\label{eq4}
\end{eqnarray}
\noindent
where $k^{(n)}(i\xi,k_{\bot})$ is defined as
\begin{equation}
k^{(n)}(i\xi,k_{\bot})=\left[k_{\bot}^2+
\varepsilon^{(n)}(i\xi)\mu^{(n)}(i\xi)
\frac{\xi^2}{c^2}\right]^{1/2}
\label{eq5}
\end{equation}
\noindent
and $\varepsilon^{(0)}(i\xi)=\mu^{(0)}(i\xi)=1$ are the dielectric
permittivity and magnetic permeability of vacuum.

In this section we restrict ourselves by the interaction of a
microparticle with an isolated film made of magnetodielectric
material. In this case
$\varepsilon^{(2)}(i\xi)=\mu^{(2)}(i\xi)=1$.
For a dielectric film there exist finite limiting values
$\varepsilon^{(1)}(0)\equiv\varepsilon_0^{(1)}$ and
$\mu^{(1)}(0)\equiv\mu_0^{(1)}$ for
the dielectric permittivity and magnetic permeability of film
material, respectively. Using Eq.~(\ref{eq5}), this leads to
$k^{(n)}(0,k_{\bot})=k_{\bot}$. It is convenient to perform all
subsequent calculations in terms of the dimensionless variable
$y=2ak_{\bot}$. Then, from Eq.~(\ref{eq4}) one obtains
\begin{eqnarray}
&&
r_{\rm TM}^{(0,1)}(0,k_{\bot})=-r_{\rm TM}^{(1,2)}(0,k_{\bot})
=\frac{\varepsilon_0^{(1)}-1}{\varepsilon_0^{(1)}+1}
\equiv r_{\varepsilon}^{(1)},
\nonumber \\[-2mm]
&&
\label{eq6}\\[-1mm]
&&
r_{\rm TE}^{(0,1)}(0,k_{\bot})=-r_{\rm TE}^{(1,2)}(0,k_{\bot})
=\frac{\mu_0^{(1)}-1}{\mu_0^{(1)}+1}
\equiv r_{\mu}^{(1)},
\nonumber
\end{eqnarray}
\noindent
and from Eq.~(\ref{eq3}) the reflection coefficients
$R_{\rm TM,TE}$ at zero frequency are found
\begin{eqnarray}
&&
R_{\rm TM}(0,y)=
\frac{r_{\varepsilon}^{(1)}\left(1-e^{-yD/a}\right)}{1-
{r_{\varepsilon}^{(1)}}^2e^{-yD/a}},
\nonumber \\[-2mm]
&&
\label{eq7}\\[-1mm]
&&
R_{\rm TE}(0,y)=
\frac{r_{\mu}^{(1)}\left(1-e^{-yD/a}\right)}{1-
{r_{\mu}^{(1)}}^2e^{-yD/a}}.
\nonumber
\end{eqnarray}

The Casamir-Polder free energy can be calculated by
Eq.~(\ref{eq2}).
In terms of the variable $y$ it takes the form
\begin{eqnarray}
&&
{\cal F}(a,T)=-\frac{k_BT}{8a^3}\int_{0}^{\infty}\!\!
y^2dye^{-y}
\label{eq8} \\
&&~~~~~~~~~~\times
[\alpha(0)R_{\rm TM}(0,y)+
\beta(0)R_{\rm TE}(0,y)].
\nonumber
\end{eqnarray}
\noindent
The main contribution to this integral is given by
$y\sim 1$. Then, we assume that the film thickness is
much smaller then the separation distance to a microparticle,
$D\ll a$, and expand in Eq.~(\ref{eq7}) in powers of
small parameter $D/a$
\begin{eqnarray}
&&
R_{\rm TM}(0,y)\approx
\frac{{\varepsilon_{0}^{(1)}}^2-1}{4
\varepsilon_{0}^{(1)}}\,y\,\frac{D}{a},
\nonumber \\[-2mm]
&&
\label{eq9}\\[-1mm]
&&
R_{\rm TE}(0,y)\approx
\frac{{\mu_{0}^{(1)}}^2-1}{4
\mu_{0}^{(1)}}\,y\,\frac{D}{a}.
\nonumber
\end{eqnarray}
\noindent
Substituting Eq.~(\ref{eq9}) in Eq.~(\ref{eq8}) and
integrating, we arrive at
\begin{equation}
{\cal F}(a,T)\approx
-\frac{3k_BTD}{16a^4}
\left[\alpha(0)\frac{{\varepsilon_{0}^{(1)}}^2-1}{\varepsilon_{0}^{(1)}}+
\beta(0)\frac{{\mu_{0}^{(1)}}^2-1}{\mu_{0}^{(1)}}\right].
\label{eq10}
\end{equation}
\noindent
The respective expression for the Casimir-Polder force acting
between a particle and a thin magnetodielectric film is the
following:
\begin{equation}
{F}(a,T)\approx
-\frac{3k_BTD}{4a^5}
\left[\alpha(0)\frac{{\varepsilon_{0}^{(1)}}^2-1}{\varepsilon_{0}^{(1)}}+
\beta(0)\frac{{\mu_{0}^{(1)}}^2-1}{\mu_{0}^{(1)}}\right].
\label{eq11}
\end{equation}
\noindent
For dielectric film with no magnetic properties the Casimir-Polder
 free energy and force are obtained from Eqs.~(\ref{eq10}) and
(\ref{eq11}) by putting $\mu_0^{(1)}=1$.

It is interesting that the classical limits (\ref{eq10}) and
(\ref{eq11}) for a particle interacting with a thin film are
different from respective results for a particle interacting with
a semispace. For instance, by replacing the reflection
coefficients (\ref{eq9})
with familiar coefficients  (\ref{eq6}) describing reflection
on a semispace, the following Casimir-Polder free energy is
obtained:
\begin{equation}
{\cal F}(a,T)=
-\frac{k_BT}{4a^3}
\left[\alpha(0)\frac{{\varepsilon_{0}^{(1)}}-1}{\varepsilon_{0}^{(1)}+1}+
\beta(0)\frac{{\mu_{0}^{(1)}}-1}{\mu_{0}^{(1)}+1}\right].
\label{eq12}
\end{equation}
\noindent
It can be seen that the quantity (\ref{eq12}) decreases with
separation slower than the free energy (\ref{eq10}) calculated
for a thin film. This can be explained by the presence of new
dimensional parameter, the film thickness.
The dependences on the material parameters
$\varepsilon_{0}^{(1)},\,\mu_{0}^{(1)}$ in Eqs.~(\ref{eq10})
and (\ref{eq12}) are also different.

\section{Microparticle interacting with a metallic film}

Now we consider the classical Casimir-Polder interaction of
an atom (microparticle) with a metallic film possessing
magnetic properties. In this case
$|\varepsilon^{(1)}(\omega)|\to\infty$ when $\omega\to 0$.
Because of this, some care is needed when obtaining the
zero-frequency values in Eqs.~(\ref{eq4})
and (\ref{eq5}). At low frequencies the dielectric permittivity
of metals behaves as
$\varepsilon^{(1)}(i\xi)\approx{\omega_p^{(1)}}^2/(\xi\gamma)$,
where $\omega_p^{(1)}$ is the plasma frequency and $\gamma$
is the relaxation parameter of film metal.
It is important to underline that the Drude model takes into
account the relaxation properties of free electrons.
At high frequencies, much larger than the relaxation parameter,
one can neglect by the relaxation processes. Then the
dielectric permittivity of metals behaves at high frequencies
as $\varepsilon^{(1)}(i\xi)\approx{\omega_p^{(1)}}^2/\xi^2$
in acordance to the plasma model.
Surprisingly, the experimental data of several experiments on
measuring the Casimir force between metallic test bodies
\cite{49} exclude the Lifshitz theory using the Drude model and
are consistent with the plasma model behavior extrapolated from
high frequencies down to zero frequency. For this reason
below we use both models when they lead to dissimilar
results for the classical Casimir-Polder free energy and force.

As is seen in Eq.~(\ref{eq5}), for both models of
$\varepsilon^{(1)}(i\xi)$ the quantity $k^{(1)}(0,k_{\bot})$
takes finite values. Then, from Eqs.~(\ref{eq3}) and
(\ref{eq4}), independently of the model used, one obtains
\begin{eqnarray}
&&
r_{\rm TM}^{(0,1)}(0,k_{\bot})=
-r_{\rm TM}^{(1,2)}(0,k_{\bot})=1,
\nonumber \\
&&
R_{\rm TM}(0,k_{\bot})=1.
\label{eq13}
\end{eqnarray}

Let us assume now that the low-frequency behavior of the film
metal is described by the Drude model. In this case from
Eq.~(\ref{eq5}) one obtains $k^{(1)}(0,k_{\bot})=k_{\bot}$ and
the reflection coefficients $r_{\rm TE}^{(0,1)}(0,k_{\bot})$
and $r_{\rm TE}^{(1,2)}(0,k_{\bot})$ are again given by the
second line in Eq.~(\ref{eq6}). Substituting them in
Eq.~(\ref{eq3}) and using the variable $y$, we find the same
coefficient $R_{\rm TE}(0,y)$, as the one presented in the second
line in Eq.~(\ref{eq7}). We expand it in powers of a small
parameter $D/a$ [see the second formula in Eq.~(\ref{eq9})]
and substitute to Eq.~(\ref{eq8}) together with the second
equality in Eq.~(\ref{eq13}). The result is
\begin{equation}
{\cal F}_D(a,T)\approx
-\frac{k_BT}{4a^3}
\left[\alpha(0)+
3\beta(0)\frac{{\mu_{0}^{(1)}}^2-1}{4\mu_{0}^{(1)}}\,
\frac{D}{a}\right].
\label{eq14}
\end{equation}
\noindent
The respective Casimir-Polder force is given by
\begin{equation}
{F}_D(a,T)\approx
-\frac{3k_BT}{4a^4}
\left[\alpha(0)+
\beta(0)\frac{{\mu_{0}^{(1)}}^2-1}{\mu_{0}^{(1)}}\,
\frac{D}{a}\right].
\label{eq15}
\end{equation}
\noindent
These results are different from the case of a microparticle
interacting with a semispace made of magnetic metal.
For example, the free energy of the classical
Casimir-Polder interaction between a microparticle and a semispace
is given by
\begin{equation}
{\cal F}(a,T)=
-\frac{k_BT}{4a^3}
\left[\alpha(0)+
\beta(0)\frac{{\mu_{0}^{(1)}}-1}{\mu_{0}^{(1)}+1}\right].
\label{eq16}
\end{equation}
\noindent
This can be obtained from Eq.~(\ref{eq12}) in the limiting case
$\varepsilon_0^{(1)}\to\infty$.

As is seen from the comparison of Eq.~(\ref{eq14}) with
Eq.~(\ref{eq16}), for metallic films described by the Drude model
the film thickness does not influence the Casimir-Polder free
energy of only electrically polarizable atoms (microparticles)
or of nonmagnetic metals. At the same time, for a magnetizable
atom (microparticle) interacting with thin film made of
magnetic metal the respective contribution to the Casimir-Polder
free energy in Eq.~(\ref{eq14}) depends on the film thickness
and demonstrates an alternative dependence on separation, as
compared with the case of a semispace in Eq.~(\ref{eq16}).

We continue by considering an atom (microparticle) interacting
with metallic film whose low-frequency response is described
by the plasma model. In this case the contribution of the TE
mode remains the same, as for the Drude model, i.e., is given
by Eq.~(\ref{eq13}). The contribution of the TE mode is,
however, different. From Eqs.~(\ref{eq4}) and (\ref{eq5}),
using the variable $y$, one finds
\begin{equation}
r_{\rm TE}^{(0,1)}(0,y)=
-r_{\rm TE}^{(1,2)}(0,y)=\frac{\mu_0^{(1)}y-
\sqrt{y^2+\mu_0^{(1)}{{\tilde{\omega}}_p^{(1)}}
{\vphantom{{\tilde{\omega}}_p^{(1)}}}^2}}{\mu_0^{(1)}y+
\sqrt{y^2+\mu_0^{(1)}{{\tilde{\omega}}_p^{(1)}}
{\vphantom{{\tilde{\omega}}_p^{(1)}}}^2}},
\label{eq17}
\end{equation}
\noindent
where
\begin{equation}
\tilde{\omega}_p^{(1)}=\frac{\omega_p^{(1)}}{\omega_c}
\equiv\frac{2a\omega_p^{(1)}}{c}=
\frac{2a}{\delta^{(1)}},
\label{eq18}
\end{equation}
\noindent
and $\delta^{(1)}$ is the penetration depth of the
electromagnetic oscillations into the film material.
Then Eq.~(\ref{eq17}) can be rewritten as
\begin{equation}
r_{\rm TE}^{(0,1)}(0,y)=
\frac{\frac{\delta^{(1)}\sqrt{\mu_0^{(1)}}}{2a}y-
\left[1+\left(\frac{\delta^{(1)}\sqrt{\mu_0^{(1)}}}{2a}\right)^2\!
\frac{y^2}{{\mu_0^{(1)}}^2}\right]^{1/2}}{\frac{\delta^{(1)}
\sqrt{\mu_0^{(1)}}}{2a}y+
\left[1+\left(\frac{\delta^{(1)}\sqrt{\mu_0^{(1)}}}{2a}\right)^2\!
\frac{y^2}{{\mu_0^{(1)}}^2}\right]^{1/2}}.
\label{eq19}
\end{equation}

Now we take into account that even for rather bad metals the
penetration depth in not larger than $\delta^{(1)}\approx
100\,$nm.
Keeping in mind that the classical limit holds at
$a>5\mu\mbox{m}$, one concludes that
$\delta^{(1)}/(2a)\lesssim 10^{-2}$. Then for typical values
of $\mu_0^{(1)}\lesssim 100$ we can use the small parameter
\begin{equation}
\frac{\delta^{(1)}}{2a}\sqrt{\mu_0^{(1)}}\ll 1.
\label{eq20}
\end{equation}
Expanding Eq.~(\ref{eq19}) in powers of this parameter for
$y\sim 1$, one obtains
\begin{equation}
r_{\rm TE}^{(0,1)}(0,y)\approx
-1+\frac{\delta^{(1)}}{a}\sqrt{\mu_0^{(1)}}y.
\label{eq21}
\end{equation}

{}From Eq.~(\ref{eq3}) we now find
\begin{equation}
R_{\rm TE}(0,y)=\frac{r_{\rm TE}^{(0,1)}(0,y)\left[1-
e^{-\frac{D}{a}(y^2+\mu_0^{(1)}{{\tilde{\omega}}_p^{(1)}}
{\vphantom{{\tilde{\omega}}_p^{(1)}}}^2)^{1/2}}\right]}{1-
{r_{\rm TE}^{(0,1)}}^2(0,y)
e^{-\frac{D}{a}(y^2+\mu_0^{(1)}{{\tilde{\omega}}_p^{(1)}}
{\vphantom{{\tilde{\omega}}_p^{(1)}}}^2)^{1/2}}},
\label{eq22}
\end{equation}
\noindent
where $r_{\rm TE}^{(0,1)}(0,y)$ is defined in Eq.~(\ref{eq21}).
Note that using Eq.~(\ref{eq20}) the power of the exponent in
Eq.~(\ref{eq22}) can be approximately written as
\begin{equation}
-\frac{D}{a}(y^2+\mu_0^{(1)}{{\tilde{\omega}}_p^{(1)}}
{\vphantom{{\tilde{\omega}}_p^{(1)}}}^2)^{1/2}\approx
-2\sqrt{\mu_0^{(1)}}\frac{D}{\delta^{(1)}}.
\label{eq23}
\end{equation}

Further consideration depends on the relationship between the
film parameters and separation distance.
In the first case the inequality holds
\begin{equation}
\frac{D}{\delta^{(1)}}\gg\frac{\delta^{(1)}}{a}.
\label{eq24}
\end{equation}
\noindent
This is always valid at sufficiently large separations.
Then, expanding Eq.~(\ref{eq22}) in powers of the small parameter
(\ref{eq20}) with account of Eq.~(\ref{eq23}), one obtains
\begin{equation}
R_{\rm TE}(0,y)\approx -1+
\frac{\delta^{(1)}}{a}\sqrt{\mu_0^{(1)}}y
\coth\frac{D\sqrt{\mu_0^{(1)}}}{\delta^{(1)}}.
\label{eq25}
\end{equation}
\noindent
Substituting Eqs.~(\ref{eq13}) and (\ref{eq25}) in Eq.~(\ref{eq8})
 and performing integration with respect to $y$, we arrive at
\begin{eqnarray}
&&
{\cal F}_p(a,T)\approx
-\frac{k_BT}{4a^3}
\left[
\vphantom{\frac{\delta^{(1)}}{a}\sqrt{\mu_0^{(1)}}
\coth\frac{D\sqrt{\mu_0^{(1)}}}{\delta^{(1)}}}
\alpha(0)-\beta(0)\right.
\label{eq26} \\
&&~~~~~~+\left.
3\beta(0)\frac{\delta^{(1)}}{a}\sqrt{\mu_0^{(1)}}
\coth\frac{D\sqrt{\mu_0^{(1)}}}{\delta^{(1)}}\right].
\nonumber
\end{eqnarray}
\noindent
The respective result for the Casimir-Polder force is
\begin{eqnarray}
&&
{F}_p(a,T)=
-\frac{3k_BT}{4a^4}
\left[
\vphantom{\frac{\delta^{(1)}}{a}\sqrt{\mu_0^{(1)}}
\coth\frac{D\sqrt{\mu_0^{(1)}}}{\delta^{(1)}}}
\alpha(0)-\beta(0)\right.
\label{eq27} \\
&&~~~~~~\left.+
4\beta(0)\frac{\delta^{(1)}}{a}\sqrt{\mu_0^{(1)}}
\coth\frac{D\sqrt{\mu_0^{(1)}}}{\delta^{(1)}}\right].
\nonumber
\end{eqnarray}
\noindent
By comparing Eqs.~(\ref{eq26}) and (\ref{eq27}) with
respective Eqs.~(\ref{eq14}) and (\ref{eq15}) obtained for
films described by the Drude model, one can see that for
only electrically polarizable atoms the results are
coinciding. However, for magnetizable microparticles, i.e.,
for $\beta(0)\neq 0$, the plasma model approach leads to
different predictions for the Casimir-Polder free energy
and force even if the film is made of a nonmagnetic metal.

In the second case the inequality opposite to Eq.~(\ref{eq21})
holds, i.e.,
\begin{equation}
\frac{D}{\delta^{(1)}}\ll\frac{\delta^{(1)}}{a}.
\label{eq28}
\end{equation}
\noindent
This is the case of very thin films and not too large
separation distances. Expanding Eq.~(\ref{eq22}) in powers of
the small parameter $\sqrt{\mu_0^{(1)}}D/\delta^{(1)}$, we
obtain the result
\begin{equation}
R_{\rm TE}(0,y)\approx -
\frac{aD}{aD+{\delta^{(1)}}^2y},
\label{eq29}
\end{equation}
\noindent
which does not depend on $\mu_0^{(1)}$. Substituting
Eqs.~(\ref{eq13}) and (\ref{eq29}) in Eq.~(\ref{eq8})
and integrating, one arrives at
\begin{eqnarray}
&&
{\cal F}_p(a,T)\approx
-\frac{k_BT}{8a^3}
\left\{
\vphantom{\left[\left(\frac{aD}{{\delta^{(1)}}^2}
\right)^2\right]}
2\alpha(0)-\beta(0)\frac{aD}{{\delta^{(1)}}^2}\right.
\label{eq30} \\
&&~~~~\times\left.
\left[1-\frac{aD}{{\delta^{(1)}}^2}+
\left(\frac{aD}{{\delta^{(1)}}^2}\right)^2
e^{{aD}/{{\delta^{(1)}}^2}}
\Gamma\left(0,\frac{aD}{{\delta^{(1)}}^2}\right)\right]
\right\},
\nonumber
\end{eqnarray}
\noindent
where $\Gamma(x,y)$ is the incomplete gamma function.
Then, expanding in powers of the small parameter
${aD}/{\delta^{(1)}}^2$, we finally obtain
\begin{equation}
{\cal F}_p(a,T)\approx
-\frac{k_BT}{4a^3}
\left[\alpha(0)-\beta(0)\frac{aD}{2{\delta^{(1)}}^2}\right]
\label{eq31}
\end{equation}
\noindent
and the respective expression for the Casimir-Polder force
\begin{equation}
{F}_p(a,T)\approx
-\frac{3k_BT}{4a^4}
\left[\alpha(0)-\beta(0)\frac{aD}{3{\delta^{(1)}}^2}\right].
\label{eq32}
\end{equation}
\noindent
Note that for films satisfying Eq.~(\ref{eq28}) main
contributions to the classical Casimir-Polder free energy
and force coincide with those in Eqs.~(\ref{eq14}) and
(\ref{eq15}) derived using the Drude model.

For a microparticle above a thick metallic plate (semispace)
described by the plasma model the Casimir-Polder free energy
is given by
\begin{equation}
{\cal F}_p(a,T)\approx
-\frac{k_BT}{4a^3}
\left[\alpha(0)-\beta(0)\left(1-3
\frac{\delta^{(1)}\sqrt{\mu_0^{(1)}}}{a}
\right)\right].
\label{eq33}
\end{equation}
This is in agreement with the case of a metallic film under
the condition (\ref{eq24}), where, in addition, the film
thickness $D\gg\delta^{(1)}$.
\noindent

\section{Microparticle interacting with a graphene sheet}

Here we consider the classical Casimir-Polder interaction of
an atom (microparticle) with a graphene sheet and compare the
obtained results with the above results found for thin
material films. We describe graphene in the framework of the
Dirac model \cite{24}. The reflection coefficients of the
electromagnetic oscillations on a graphene sheet at zero
Matsubara frequency in the limit of high temperature
(large separation) are given by \cite{38,40,50}
\begin{eqnarray}
&&
R_{\rm TM}(0,y)\approx 1-\frac{2y}{\tilde{\Pi}_{00}(0)},
\label{eq34} \\
&&
R_{\rm TE}(0,y)\approx -
\frac{e^2y}{3a\Delta}\left(\frac{v_F}{c}\right)^2
\tanh\frac{\Delta}{2k_BT}.
\nonumber
\end{eqnarray}
\noindent
Here, $v_F\approx c/300$ is the Fermi velocity,
$e$ is the electron charge, $\Delta\leq 0.1\,$eV is the
energy gap parameter (for a pristine graphene $\Delta=0$),
and $\tilde{\Pi}_{00}(0)$ is the 00-component of the
dimensionless polarization tensor calculated at zero
frequency. It is given by \cite{40,50}
\begin{equation}
\tilde{\Pi}_{00}(0)\approx
\frac{32e^2ak_BT}{\hbar^2v_F^2}
\ln\left(2\cosh\frac{\Delta}{2k_BT}\right).
\label{eq35}
\end{equation}

Substituting Eqs.~(\ref{eq34}) and (\ref{eq35}) in
Eq.~(\ref{eq8}), we obtain the classical free energy for the
Casimir-Polder interaction of a microparticle with a graphene
sheet
\begin{eqnarray}
&&
{\cal F}(a,T)\approx -\frac{k_BT}{4a^3}\left\{\alpha(0)
\left[1-
\frac{3\hbar^2v_F^2}{16e^2k_BTa
\ln\left(2\cosh\frac{\Delta}{2k_BT}\right)}\right]
\right.
\nonumber \\
&&~~~~~~~~
\left.
-\beta(0)\frac{e^2}{a\Delta}\left(\frac{v_F}{c}\right)^2
\tanh\frac{\Delta}{2k_BT}
\vphantom{\frac{3\hbar^2v_F^2}{16e^2k_BTa
\ln\left(\cosh\frac{\Delta}{2k_BT}\right)}}
\right\}.
\label{eq36}
\end{eqnarray}
\noindent
In the limiting case of a pristine (gapless) graphene
($\Delta=0$), Eq.~(\ref{eq36}) results in
\begin{eqnarray}
&&
{\cal F}(a,T)\approx -\frac{k_BT}{4a^3}\left\{\alpha(0)
\left[1-
\frac{3\hbar^2v_F^2}{16\ln\!{2}\,e^2k_BTa}\right]
\right.
\nonumber \\
&&~~~~~~~~
\left.
-\beta(0)\frac{e^2}{2k_BTa}\left(\frac{v_F}{c}\right)^2
\right\}.
\label{eq37}
\end{eqnarray}
\noindent
The respective expressions for the classical Casimir-Polder force
in the case of graphene with $\Delta\neq 0$ and $\Delta=0$ are
\begin{eqnarray}
&&
{F}(a,T)\approx -\frac{3k_BT}{4a^4}\left\{\alpha(0)
\left[1-
\frac{\hbar^2v_F^2}{4e^2k_BTa
\ln\left(2\cosh\frac{\Delta}{2k_BT}\right)}\right]
\right.
\nonumber \\
&&~~~~~~~~
\left.
-\beta(0)\frac{4e^2}{3a\Delta}\left(\frac{v_F}{c}\right)^2
\tanh\frac{\Delta}{2k_BT}
\vphantom{\frac{3\hbar^2v_F^2}{16e^2k_BTa
\ln\left(\cosh\frac{\Delta}{2k_BT}\right)}}
\right\},
\label{eq38}\\
&&
{F}(a,T)\approx -\frac{3k_BT}{4a^4}\left\{\alpha(0)
\left[1-
\frac{\hbar^2v_F^2}{4\ln\!{2}\,e^2k_BTa}\right]
\right.
\nonumber \\
&&~~~~~~~~
\left.
-\beta(0)\frac{2e^2}{3k_BTa}\left(\frac{v_F}{c}\right)^2
\right\}.
\nonumber
\end{eqnarray}
\noindent
Note that main contributions to Eqs.~(\ref{eq36})--(\ref{eq38})
coincide with those in Eqs.~(\ref{eq14}) and (\ref{eq15})
for a microparticle
interacting with a metallic film described by the Drude model
or in Eqs.~(\ref{eq31}) and (\ref{eq32}) derived for the case
when the film metal is described by the plasma model under
a condition (\ref{eq28}).

{}From Eqs.~(\ref{eq36})--(\ref{eq38}) it is seen that main
contributions to the Casimir-Polder free energy and force
due to electric atomic polarizability and contributions due
to magnetic atomic polarizability do not depend on the Planck
constant, as it should be in the classical limit.
Furthermore, all contributions due
to magnetic atomic polarizability are suppressed by the small
parameter $(v_F/c)^2$.

It is interesting to find the application region of our
asymptotic results in Eqs.~(\ref{eq34})--(\ref{eq38}).
This aim consists of two parts. Numerical computations using
the exact expressions for the polarization tensor \cite{44}
show that for a pristine graphene ($\Delta=0$)
Eqs.~(\ref{eq34}) and (\ref{eq35}) correctly reproduce the
zero-frequency contribution to the Lifshitz formula at
separations $a>150\,$nm. For a graphene sheet with
$\Delta=0.1\,$eV the same is correct at $a>500\,$nm.

Then one should find starting from what separation all
other contributions to the Lifshitz formula can be neglected,
so that the total result is given by the zero-frequency
contribution. It is well known \cite{37,38,40,41} that for
two graphene sheets and for a graphene sheet interacting
with a plate made of an ordinary material the classical limit
starts at much shorter separations than for two material
plates. In Ref.~\cite{44} the total Casimir-Polder free
energy of an electrically polarizable atom interacting with
graphene sheet with various $\Delta$ is computed at
$T=300\,$K as a function of separation over the region from
50\,nm to $6\,\mu$m.
By comparing these computational results with our analytic
expressions (\ref{eq36}) and (\ref{eq37}) one can conclude
that at separations $a>1.5\,\mu$m the zero-frequency term
of the Lifshitz formula comtributes more than 98\% of the
total Casimir-Polder free energy independently of the value
of $\Delta\leq 0.1\,$eV. Thus, for an atom-graphene interaction
the classical limit is achieved at larger separation distances
than for graphene-graphene interactions (in the latter case it
is achieved at approximately 300\,nm). Recall that for two plates made
of an ordinary material or for an atom interacting with such a
plate the classical limit is achieved at
separations above $5\,\mu$m. In this regard the configuration
of a microparticle above a graphene sheet is somewhat
intermediate between the configurations of two graphene sheets
and two material plates or an atom above a plate.

In Ref.~\cite{45} it was found that at zero temperature the
Casimir-Polder interaction between an atom and graphene is
approximately 5\% of the same interaction between an atom and
an ideal metal plane. Here, we compare the Casimir-Polder
interactions between an atom and graphene and  between an atom
and an ideal metal plane at room temperature. The solid line
in Fig.~2 shows the ratio of atom-graphene free energy
${\cal F}$, computed using the exact expression for the
polarization tensor with account of all contributing Matsubara
frequencies in Ref.~\cite{44}, to the free energy
${\cal F}_{\rm im}$ of the same atom interacting with
an ideal-metal plane.
Both free energies are computed at $T=300\,$K
as a function of separation in the region from 200\,nm to
$6\,\mu$m. Note that within the separation region considered
only the static electric polarizability gives the dominant
contribution to atom-graphene free energy. Because of this,
we use the free energy of an atom-ideal plane interaction under
the assumption that this atom is characterized by only the
static electric polarizability $\alpha(0)$. This quantity
is given by \cite{11}
\begin{eqnarray}
&&
{\cal F}_{\rm im}(a,T)=-\frac{k_BT\alpha(0)}{4a^3}
\left[1+\frac{2}{e^{\tau}-1}+\frac{2\tau e^{\tau}}{(e^{\tau}-1)^2}
\right.
\nonumber \\
&&~~~~~~~~~~~~
\left.
+\frac{\tau^2 e^{\tau}(e^{\tau}+1)}{(e^{\tau}-1)^3}\right],
\label{eq39}
\end{eqnarray}
\noindent
where $\tau\equiv 4\pi ak_BT/(\hbar c)$. The value of $\Delta$
in the region from 0 to 0.1\,eV influences the computational
results for the free energy of atom-graphene interaction
only in the fourth significant figure.

As can be seen in Fig.~2, at $a=200\,$nm the free energy of
atom-graphene interaction is equal to approximately 5\% of the
free energy of atom interacting with ideal metal plane, as was
found in Ref.~\cite{45} at zero temperature. However, with
increasing separation the ratio ${\cal F}/{\cal F}_{\rm im}$
quickly increases. Thus, at separations 1 and $5\,\mu$m it is
equal to 27\% and 98\%, respectively. At $a\geq 6\,\mu$m
the Casimir-Polder free energy and force for atom-graphene
interaction become equal to those for atom interacting with an
ideal-metal plane. Because of this, the finding of
Ref.~\cite{45}, where the 5\% ratio was prolonged up to
$5\,\mu$m, are applicable to only zero temperature and cannot
be extrapolated to the case of room temperature.

In the end of this section we briefly discuss the impact of
dynamic atomic polarizability and nonzero penetration depth of
the electromagnetic oscillations into real metal on the
conclusions obtained. If one considers an atom described by the
dynamic electric polarizability and an Au plate characterized
by the frequency-dependent dielectric permittivity, the
magnitudes of the resulting Casimir-Polder free energy are
equal \cite{10} to $\approx 50$\% and $\approx 90$\% of the
respective magnitudes calculated by using the static
polarizability
of an atom and an ideal-metal plane at separations 200\,nm and
$1\,\mu$m, respectively. Then one can conclude that the
magnitudes of the Casimir-Polder free energy between a real atom
and a graphene sheet are equal to approximately 10\% and 30\% of
the magnitudes calculated for the same atom interacting with
an Au plate at the same respective separations $a=200\,$nm and
$a=1\,\mu$m.

\section{Microparticle interacting with thin film deposited on
a substrate}

Now we consider the classical Casimir-Polder interaction of an
atom (microparticle) with thin material film of thickness $D$
deposited on thick substrate (semispace). We first assume that
both the film and the substrate are made of magnetodielectric
materials  with finite $\varepsilon_0^{(1)}$ and
$\varepsilon_0^{(2)}$ (see Fig.~1).
{}From Eqs.~(\ref{eq4}) and (\ref{eq5}) one obtains
\begin{equation}
r_{\rm TM}^{(0,1)}(0,y)=r_{\varepsilon}^{(1)}, \quad
r_{\rm TM}^{(1,2)}(0,y)=
\frac{\varepsilon_0^{(2)}-
\varepsilon_0^{(1)}}{\varepsilon_0^{(2)}+\varepsilon_0^{(1)}}
\equiv r_{\varepsilon}^{(2,1)},
\label{eq40}
\end{equation}
\noindent
where $r_{\varepsilon}^{(1)}$ is defined in Eq.~(\ref{eq6}).
Substituting Eq.~(\ref{eq40}) in Eq.~(\ref{eq3})
for $R_{\rm TM}$, we find
\begin{equation}
R_{\rm TM}(0,y)=\frac{r_{\varepsilon}^{(1)}+
r_{\varepsilon}^{(2,1)}e^{-Dy/a}}{1+r_{\varepsilon}^{(1)}
r_{\varepsilon}^{(2,1)}e^{-Dy/a}}\, .
\label{eq41}
\end{equation}
\noindent
Expanding here in powers of a small parameter $D/a$, we
arrive at
\begin{equation}
R_{\rm TM}(0,y)\approx
\frac{\varepsilon_0^{(2)}-1}{\varepsilon_0^{(2)}+1}-
\frac{{\varepsilon_0^{(2)}}^2-
{\varepsilon_0^{(1)}}^2}{\varepsilon_0^{(1)}(\varepsilon_0^{(2)}+1)^2}
\,\frac{D}{a}\, y.
\label{eq42}
\end{equation}
\noindent
The respective result for $R_{\rm TE}(0,y)$ is obtained from
Eq.~(\ref{eq42}) by the replacements
$\varepsilon_0^{(1)}\to\mu_0^{(1)}$ and
$\varepsilon_0^{(2)}\to\mu_0^{(2)}$.

Substituting $R_{\rm TM}$ from Eq.~(\ref{eq42}) and
$R_{\rm TE}$ to Eq.~(\ref{eq18}), after the integration with
respect to $y$ we obtain the classical Casimir-Polder free
energy
\begin{eqnarray}
&&
{\cal F}(a,T)\approx -\frac{k_BT}{4a^3}\left\{
\alpha(0)\frac{\varepsilon_0^{(2)}-1}{\varepsilon_0^{(2)}+1}
\left[1-3\frac{{\varepsilon_0^{(2)}}^2-
{\varepsilon_0^{(1)}}^2}{\varepsilon_0^{(1)}({\varepsilon_0^{(2)}}^2-1)}
\,\frac{D}{a}
\right]\right.
\nonumber \\
&&~~~~~
\left.
+\beta(0)\frac{\mu_0^{(2)}-1}{\mu_0^{(2)}+1}
\left[1-3\frac{{\mu_0^{(2)}}^2-
{\mu_0^{(1)}}^2}{\mu_0^{(1)}({\mu_0^{(2)}}^2-1)}
\,\frac{D}{a}
\right]\right\}.
\label{eq43}
\end{eqnarray}
\noindent
The respective classical Casimir-Polder force between a
microparticle and thin material film deposited on a substrate
is given by
\begin{eqnarray}
&&
{F}(a,T)\approx -\frac{3k_BT}{4a^4}\left\{
\alpha(0)\frac{\varepsilon_0^{(2)}-1}{\varepsilon_0^{(2)}+1}
\left[1-4\frac{{\varepsilon_0^{(2)}}^2-
{\varepsilon_0^{(1)}}^2}{\varepsilon_0^{(1)}({\varepsilon_0^{(2)}}^2-1)}
\,\frac{D}{a}
\right]\right.
\nonumber \\
&&~~~~~
\left.
+\beta(0)\frac{\mu_0^{(2)}-1}{\mu_0^{(2)}+1}
\left[1-4\frac{{\mu_0^{(2)}}^2-
{\mu_0^{(1)}}^2}{\mu_0^{(1)}({\mu_0^{(2)}}^2-1)}
\,\frac{D}{a}
\right]\right\}.
\label{eq44}
\end{eqnarray}

It is of interest to consider specific cases of Eqs.~(\ref{eq43})
and (\ref{eq44}). Thus, for a nonmagnetic film ($\mu_0^{(1)}=1$) on
a magnetic substrate we have from Eq.~(\ref{eq43})
\begin{eqnarray}
&&
{\cal F}(a,T)\approx -\frac{k_BT}{4a^3}\left\{
\alpha(0)\frac{\varepsilon_0^{(2)}-1}{\varepsilon_0^{(2)}+1}
\left[1-3\frac{{\varepsilon_0^{(2)}}^2-
{\varepsilon_0^{(1)}}^2}{\varepsilon_0^{(1)}({\varepsilon_0^{(2)}}^2-1)}
\,\frac{D}{a}
\right]\right.
\nonumber \\
&&~~~~~
\left.
+\beta(0)\frac{\mu_0^{(2)}-1}{\mu_0^{(2)}+1}
\left(1-3\frac{D}{a}
\right)\right\},
\label{eq45}
\end{eqnarray}
\noindent
i.e., the correction term does not depend on the magnetic
permeability
of the substrate $\mu_0^{(2)}$. Similar result holds also for
the Casimir-Polder force. For an isolated film in a vacuum,
i.e., for $\varepsilon_0^{(2)}=\mu_0^{(2)}=1$,
Eqs.~(\ref{eq43}) and (\ref{eq44}) coincide with
Eqs.~(\ref{eq10}) and (\ref{eq11}), respectively.

The next case to consider is the classical Casimir-Polder
interaction between an atom (microparticle) and a dielectric
film deposited on a metallic substrate (note that in the case
of metallic film the role of substrate made of any material
is negligibly small). We start from the contribution of the TM
mode. Here the result does not depend on the used model of
substrate metal. {}From Eqs.~(\ref{eq4}) and (\ref{eq5})
one easily finds
\begin{equation}
r_{\rm TM}^{(0,1)}(0,y)=r_{\varepsilon}^{(1)}, \quad
r_{\rm TM}^{(1,2)}(0,y)=1.
\label{eq46}
\end{equation}
\noindent
Then from Eq.~(\ref{eq3})
 we obtain
\begin{equation}
R_{\rm TM}(0,y)=\frac{r_{\varepsilon}^{(1)}+
e^{-Dy/a}}{1+r_{\varepsilon}^{(1)}e^{-Dy/a}}\,.
\label{eq47}
\end{equation}
\noindent
Expanding in this equation in powers of the small parameter $D/a$,
one arrives at
\begin{equation}
R_{\rm TM}(0,y)\approx
1-\frac{1}{\varepsilon_0^{(1)}}
\,\frac{D}{a}\,y\,.
\label{eq48}
\end{equation}
\noindent
Substituting Eq.~(\ref{eq48}) in the first term on the right-hand
side of Eq.~(\ref{eq8}) and integrating, one obtains the
contribution of the TM mode to the classical Casimir-Polder free
energy
\begin{equation}
{\cal F}_{\rm TM}(a,T)\approx -\frac{k_BT\alpha(0)}{4a^3}
\left(1-\frac{3}{\varepsilon_0^{(1)}}
\,\frac{D}{a}\right).
\label{eq49}
\end{equation}

The contribution of the TE mode depends on the model of substrate
metal. We first assume that the low-frequency behavior of the
dielectric permittivity of substrate metal is described by the
Drude model.
In this case the reflection coefficient $R_{\rm TE}(0,y)$
coincides with the same coefficient for an atom interacting with
a dielectric film deposited on the dielectric substrate.
The respective contribution to the Casimir-Polder free energy is
given by the second term on the right-hand side of
Eq.~(\ref{eq43}):
\begin{equation}
{\cal F}_{\rm TE}(a,T)\approx -\frac{k_BT\beta(0)}{4a^3}
\,
\frac{\mu_0^{(2)}-1}{\mu_0^{(2)}+1}
\left[1-3\frac{{\mu_0^{(2)}}^2-
{\mu_0^{(1)}}^2}{\mu_0^{(1)}({\mu_0^{(2)}}^2-1)}
\,\frac{D}{a}
\right] .
\label{eq50}
\end{equation}
\noindent
As a result, the total classical Casimir-Polder free energy
and force in this case are given by
\begin{eqnarray}
&&
{\cal F}_{D}(a,T)\approx -\frac{k_BT}{4a^3}
\left\{\alpha(0)\left(1-\frac{3}{\varepsilon_0^{(1)}}
\,\frac{D}{a}\right)
\right.
\nonumber \\[1mm]
&&~~~~
\left.
+\beta(0)
\frac{\mu_0^{(2)}-1}{\mu_0^{(2)}+1}
\left[1-3\frac{{\mu_0^{(2)}}^2-
{\mu_0^{(1)}}^2}{\mu_0^{(1)}({\mu_0^{(2)}}^2-1)}
\,\frac{D}{a}
\right]\right\},
\nonumber \\[-2mm]
&&
\label{eq51}\\[-1mm]
&&
{F}_{D}(a,T)\approx -\frac{3k_BT}{4a^4}
\left\{\alpha(0)\left(1-\frac{4}{\varepsilon_0^{(1)}}
\,\frac{D}{a}\right)
\right.
\nonumber \\[1mm]
&&~~~~
\left.
+\beta(0)
\frac{\mu_0^{(2)}-1}{\mu_0^{(2)}+1}
\left[1-4\frac{{\mu_0^{(2)}}^2-
{\mu_0^{(1)}}^2}{\mu_0^{(1)}({\mu_0^{(2)}}^2-1)}
\,\frac{D}{a}
\right]\right\}.
\nonumber
\end{eqnarray}
For a nonmagnetic substrate these expressions are
simplified. For example, the free energy takes the form
\begin{eqnarray}
&&
{\cal F}_{D}(a,T)\approx -\frac{k_BT}{4a^3}
\left[
\vphantom{\frac{{\mu_0^{(1)}}^2-1}{\mu_0^{(1)}}}
\alpha(0)\left(1-\frac{3}{\varepsilon_0^{(1)}}
\,\frac{D}{a}\right)
\right.
\label{eq52} \\
&&~~~~
\left.
+\beta(0)\frac{3}{4}\,
\frac{{\mu_0^{(1)}}^2-1}{\mu_0^{(1)}}
\,\frac{D}{a}
\right].
\nonumber
\end{eqnarray}

Let us now assume that the metal of a substrate is described by
the plasma model. In this case the contribution of the TM mode
to the free energy is again given by Eq.~(\ref{eq49}).
The TE reflection coefficients (\ref{eq4}) are given by
\begin{eqnarray}
&&
r_{\rm TE}^{(0,1)}(0,y)=r_{\mu}^{(1)},
\label{eq53} \\
&&
r_{\rm TE}^{(1,2)}(0,y)=\frac{\mu_0^{(2)}y-\mu_0^{(1)}
\sqrt{y^2+\mu_0^{(2)}{\tilde{\omega}_p^{(2)}}
{\vphantom{{\tilde{\omega}_p^{(2)}}}}^2}}{\mu_0^{(2)}y
+\mu_0^{(1)}\sqrt{y^2+\mu_0^{(2)}{\tilde{\omega}_p^{(2)}}
{\vphantom{{\tilde{\omega}_p^{(2)}}}}^2}}\, ,
\nonumber
\end{eqnarray}
\noindent
where the normalized plasma frequency of the substrate metal
$\tilde{\omega}_p^{(2)}$ is defined in the same way as
$\tilde{\omega}_p^{(1)}$ in Eq.~(\ref{eq18}) and expressed via
the penetration depth $\delta^{(2)}$ of the electromagnetic
oscillations into the material of a substrate.
The coefficient $r_{\rm TE}^{(1,2)}$ in Eq.~(\ref{eq53})
can be identically presented as
\begin{equation}
r_{\rm TE}^{(1,2)}(0,y)=\frac{\frac{\delta^{(2)}}{2a}\,
\frac{\sqrt{\mu_0^{(2)}}}{\mu_0^{(1)}}y-\left[1+
\left(\frac{\delta^{(2)}}{2a}\right)^2
\frac{y^2}{\mu_0^{(2)}}\right]^{1/2}}{\frac{\delta^{(2)}}{2a}\,
\frac{\sqrt{\mu_0^{(2)}}}{\mu_0^{(1)}}y+\left[1+
\left(\frac{\delta^{(2)}}{2a}\right)^2
\frac{y^2}{\mu_0^{(2)}}\right]^{1/2}}.
\label{eq54}
\end{equation}
\noindent
Expanding Eq.~(\ref{eq54}) in powers of small parameter
$\delta^{(2)}\sqrt{\mu_0^{(2)}}/(2a)$, we obtain
\begin{equation}
r_{\rm TE}^{(1,2)}(0,y)\approx -1+
\frac{\delta^{(2)}}{a}\,
\frac{\sqrt{\mu_0^{(2)}}}{\mu_0^{(1)}}y
\label{eq55}
\end{equation}
\noindent
in close analogy with Eq.~(\ref{eq21}).

Substituting Eq.~(\ref{eq55}) and the quantity
$r_{\rm TE}^{(0,1)}$ from Eq.~(\ref{eq53}) in Eq.~(\ref{eq3})
and expanding in powers of small parameters
$\delta^{(2)}\sqrt{\mu_0^{(2)}}/(2a)$ and $\mu_0^{(1)}D/a$,
we find
\begin{equation}
R_{\rm TE}(0,y)\approx -1+
\sqrt{\mu_0^{(2)}}\frac{\delta^{(2)}}{a}\,y
+\mu_0^{(1)}\frac{D}{a}y.
\label{eq56}
\end{equation}
\noindent
Then from Eq.~(\ref{eq8}) one arrives to the following TE
contribution to the Casimir-Polder free energy:
\begin{equation}
{\cal F}_{\rm TE}(a,T)\approx \frac{k_BT}{4a^3}\beta(0)
\left(
1-
3\sqrt{\mu_0^{(2)}}\frac{\delta^{(2)}}{a}\,
-3\mu_0^{(1)}\frac{D}{a}\right).
\label{eq57}
\end{equation}
\noindent
Combining this with the TM contribution in Eq.~(\ref{eq49}),
we obtain the Casimir-Polder free energy of an atom interacting
with a dielectric film deposited on metallic substrate
described by the plasma metal
\begin{eqnarray}
&&
{\cal F}_{p}(a,T)\approx -\frac{k_BT}{4a^3}
\left[\alpha(0)\left(1-\frac{3}{\varepsilon_0^{(1)}}\,\frac{D}{a}
\right)\right.
\label{eq58} \\
&&
-\left.\beta(0)
\left(
1-
3\sqrt{\mu_0^{(2)}}\frac{\delta^{(2)}}{a}\,
-3\mu_0^{(1)}\frac{D}{a}\right)
\vphantom{\left(\frac{3}{\varepsilon_0^{(1)}}\right)}
\right].
\nonumber
\end{eqnarray}
\noindent
The respective expression for the Casimir-Polder force is
\begin{eqnarray}
&&
{F}_{p}(a,T)\approx -\frac{3k_BT}{4a^4}
\left[\alpha(0)\left(1-\frac{4}{\varepsilon_0^{(1)}}\,\frac{D}{a}
\right)\right.
\label{eq59} \\
&&
-\left.\beta(0)
\left(
1-
4\sqrt{\mu_0^{(2)}}\frac{\delta^{(2)}}{a}\,
-4\mu_0^{(1)}\frac{D}{a}\right)
\vphantom{\left(\frac{3}{\varepsilon_0^{(1)}}\right)}
\right].
\nonumber
\end{eqnarray}
\noindent
{}From the comparison of Eq.~(\ref{eq51}) with Eq.~(\ref{eq58})
one can conclude that the contributions to the Casimir-Polder
free energy due to atomic magnetic polarizability calculated
using the Drude and plasma models are of opposite sign. The same
is correct for the Casimir-Polder force, as it is seen from
Eqs.~(\ref{eq51}) and (\ref{eq59}).

\section{Conclusions and discussion}

In the foregoing we have derived simple analytic expressions for
the classical Casimir-Polder free energy and force for a
polarizable and magnetizable atom (microparticle) interacting
with thin films made of dielectric and metallic magnetic
materials, both isolated and deposited on substrates.
The obtained results were compared with the Casimir-Polder
interaction between an atom (microparticle) and a graphene
sheet. It was shown that the classical Casimir-Polder interaction
of an atom with a dielectric film is different from the same
interaction of an atom with a dielectric plate (semispace).
Specifically, it decreases quicker with separation and depends
on an additional dimensional parameter, the film thickness.

The classical Casimir-Polder interaction of only electrically
polarizable atoms with thin metallic films does not depend on
the model of a metal and does not depend on film thickness.
For  magnetizable atoms (microparticles) the respective
additions to the classical Casimir-Polder free energy and force
depend on the film thickness when the film metal is magnetic
and its low-frequency response
is described by the Drude model.
When the metal response is described by the plasma model, the
additions due to atomic magnetic polarizability arise for both
magnetic and nonmagnetic metals and depend on the penetration
depth of electromagnetic oscillations into the metal.
It the latter case the form of additions was shown to depend
on the relationship between the film thickness, the penetration
depth and the separation distance.

We have found analytic expressions for the
Casimir-Polder free energy and force between a polarizable and
magnetizable atom (microparticle) and a graphene sheet described
by the polarization tensor in the framework of the Dirac model.
It was shown that all contributions to these quantities due to
atomic magnetic polarizability  are suppressed, as compared to
main terms depending on atomic electric polarizability.
This conclusion is of interest for future experiments on
quantum reflection and Bose-Einstein condensation near graphene
surface. We have also shown that at room temperature the
classical limit of the Casimir-Polder interaction with graphene
is achieved at about $1.5\,\mu$m separation between an atom and
a graphene surface. This is several times as big as for two
graphene sheets, but several times as small as for two plates
made of ordinary materials or for an atom above a plate.
According to our results, at separations above $5\,\mu$m at
$T=300\,$K the Casimir-Polder interaction of atoms with graphene
is of the same strength as with an ideal-metal plane.
This differs essentially from the previously investigated case
of zero temperature where atom-graphene interaction was found
to be as much as only 5\% of the interaction of graphene with
an ideal-metal plane.
Qualitatively, the classical
Casimir-Polder interaction of atoms with
graphene was likened to the interaction with thin metallic
film.

Finally, we have obtained simple analytic expressions for
the Casimir-Polder interaction of an atom (microparticle)
with thin magnetodielectric films deposited on both
dielectric and metallic substrates made of magnetic or
nonmagnetic materials. It was shown that in the case of
magnetodielectric (dielectric) film deposited on metallic
substrate the contributions to the Csimir-Polder free energy
and force due to the magnetic atomic polarizability depend
on the used model of metal. One can conclude that the
classical Casimir-Polder interaction of atoms with thin
films and graphene discussed above differs significantly from
the interaction with thick cavity walls, and this might be
interesting for future experiments on atom-surface
interactions.


\begin{figure}[b]
\vspace*{-7cm}
\centerline{\hspace*{2cm}
\includegraphics{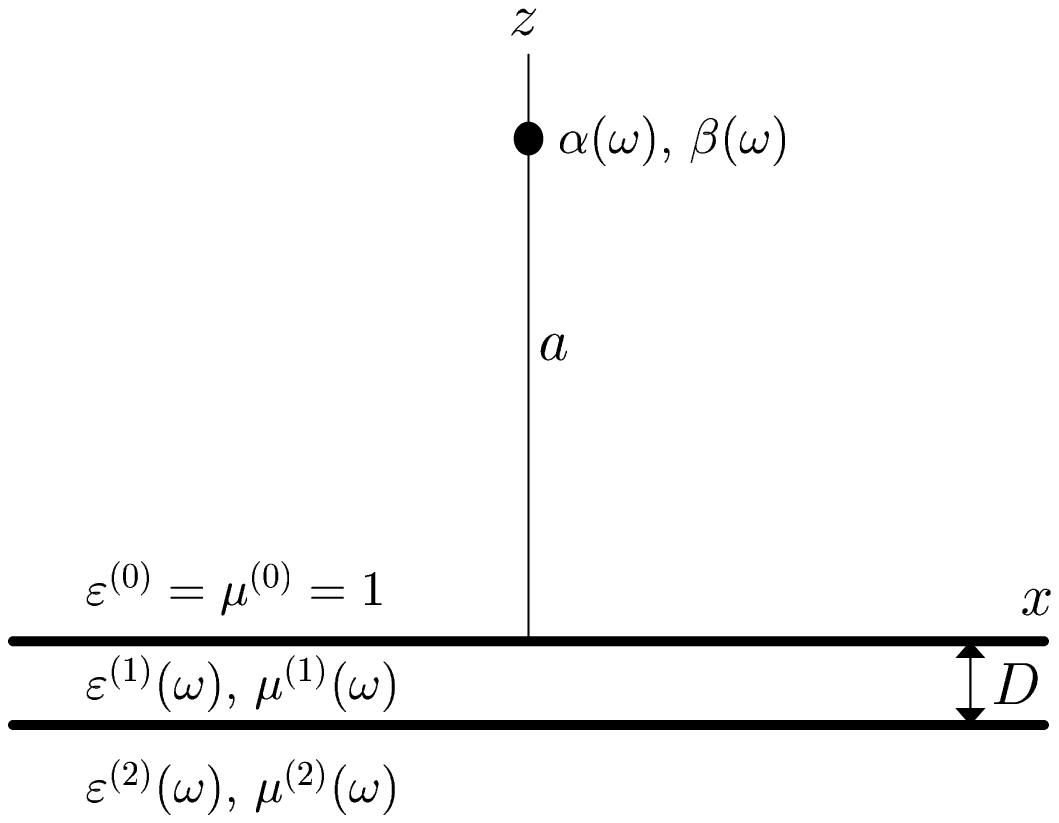}
}
\vspace*{-11cm}
\caption{\label{fg1}
The configuration of an atom (microparticle) above a thin film
deposited on a thick substrate (semispace). See text for further
discussion.
}
\end{figure}
\begin{figure}[b]
\vspace*{-7cm}
\centerline{\hspace*{2cm}
\includegraphics{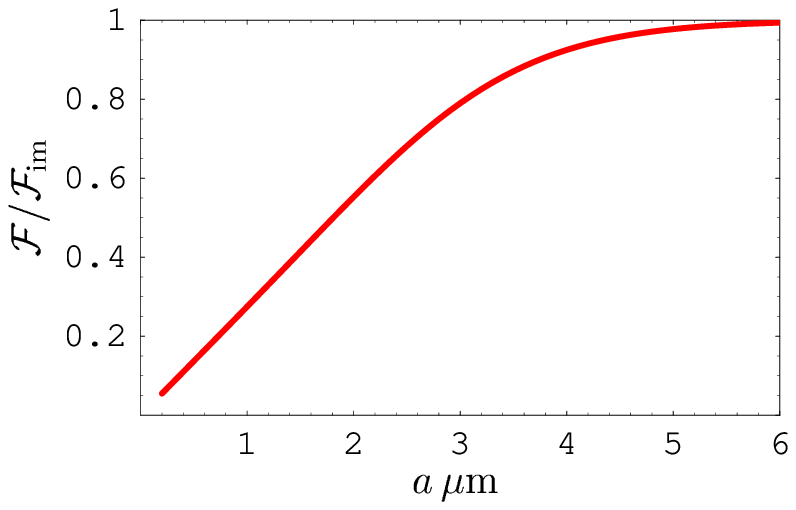}
}
\vspace*{-9.5cm}
\caption{\label{fg1}(Color online)
The free energy of the Casimir-Polder interaction of an
atom with graphene at $T=300\,$K normalized for the free
energy of the same interaction with an ideal-metal plane is
shown as a function of separation by the solid line.
}
\end{figure}
\end{document}